\documentclass{elsart}

\usepackage{natbib}

\usepackage{graphicx}

\begin{document}




\begin{frontmatter}
\title{Helical motions in the jet of  blazar 1156+295}
\author[SHAO]{X.Y.~Hong}
\author[SHAO]{D.R.~Jiang}
\author[JIVE]{L.I.~Gurvits}
\author[JIVE]{M.A.~Garrett}
\author[JIVE]{R.T.~Schilizzi}
\author[BAO]{R.D.~Nan}

\address[SHAO]{Shanghai Astronomical Observatory, 80 Nandan Road,
Shanghai, 200030, China}
\address[JIVE]{Joint Institute for VLBI in Europe, Postbus 2,
7990~AA Dwingeloo, \\
The Netherlands}
\address[BAO]{Beijing Astronomical Observatory, Datun Rd. 20A, Chaoyang District,\\
Beijing 100012, China}


\begin{abstract}

The blazar 1156+295 was observed by VLBA and EVN + MERLIN at 5 GHz in
June 1996 and February 1997 respectively. The results show that the jet
of the source has structural oscillations on the milliarcsecond scale
and turns through a large angle to the direction of the arcsecond-scale
extension. A helical jet model can explain most of the observed
properties of the radio structure in 1156+295.

\end{abstract}



\end{frontmatter}

\section{Introduction}
\label{intro}

The blazar 1156+295 ($z=0.729$, \citet{VCV}) is among the most active of
highly polarized (HPQ) and  optically violent variable (OVV) sources
\citep{G1,W1,W2} In its active phase, this source has shown
flux-density fluctuations at optical wavelengths with an amplitude of
$\sim$ 5--7\%  on time scales of 0.5 hour. Also, both the position
angle and fraction of optical polarization vary dramatically. At
cm wavelengths $\sim 10\%$ variations in total flux density
have been observed on time scales less than several days. Finally,
1156+295 has flared in $\>$ 100 MeV $\gamma$-rays at least three times
since 1992, while the quiescent $\gamma$-ray emission remains
undetected \citep{M3}.

High-resolution radio imaging has revealed arcsecond and milliarcsecond
structure consistent with the activity described above as well as with
the standard blazar model of a jet aligned nearly along the line of
sight.

On the arcsecond scale, the 1.5 GHz VLA image shows a symmetrical
structure in the north-south direction \citep{A1}. The MERLIN image at
1.6 GHz and the VLA image at 5 GHz \citep{M1} show  extended emission
on the arcsecond-scale and a knotty jet of length $\sim 2\rlap{.}''5$
at p.a. $-19^\circ$. The northern end of the jet has a hotspot, at
which point it turns through about $90^\circ $ to the east, ending in a
diffuse region of size $\sim 1\rlap{.}''5$.  There is a region of
similar diffuse emission to the south about $2''$ from the core, at
p.a. $\sim 200^{\circ}$. From the maps, 1156+295 appears to resemble a
classical double radio source seen end-on, with its northern jet
relativistically beamed towards us, and hence dominating over any
southern jet, while the diffuse emission is not beamed and appears
symmetric.

Based on VLBI observations at three frequencies, \citet{M1,M2}
estimated the apparent superluminal velocity of $26h^{-1}\,c$. This is
much larger than any proper motion reliably found for any other source
--- $\sim 2.5$ times greater than that of the next fastest object in
its redshift bin on the $\mu-z$ diagram \citep{V2}. Such a high
apparent proper motion in principle could be a projection effect:  when
the jet axis is aligned near the line of sight, small changes in the
direction of the jet axis can produce large changes in the apparent
proper motion.  Lower apparent velocities in the range of 3.5--$
8.8h^{-1}\,c$ were reported on the basis of ten epochs of geodetic
observations at 8.4 and 2.3 GHz \citep{P1}. Arguments against highly
beamed synchrotron emission from the milliarcsecond core (thus, not
necessarily a small angle between the very inner part of the jet and
viewing directions) have been reported on the basis of the  VSOP Space
VLBI observations at 1.6 GHz with the angular resolution of 4.4
$\times$ 1.4 mas \citep{HH}.

In this paper, we discuss the observations of the blazar 1156+295 at 5
GHz with the EVN + MERLIN (February 1997) in comparison with the
earlier VLBA observation. Throughout this paper, the values $H_{0}=100
h$ km\,s$^{-1}$\,Mpc$^{-1}$, $q_{0}=0.5$, and $S \propto \nu ^{\alpha}$
will be used.

\section{The observations and data reduction}
\label{features}

A full-track 12-hour EVN + MERLIN observation of the blazar 1156+295 was
carried out at 5 GHz in February 1997. The MERLIN array comprised  6
antennas (Defford, Cambridge, Knockin, Darnhall, Mark~2, and Tabley). The
central wavelength was 4994 MHz and the observing bandwidth was 14
MHz. The amplitude calibration was carried out at Jodrell Bank with
the calibrator source OQ208. The imaging was done using the NRAO
Astronomical Image Processing System (AIPS) package.

The EVN array comprised Effelsberg, WSRT, Jodrell Bank (Mark~2),
Cambridge, Onsala, Medicina, Torun, Shanghai, Urumqi, and
Hartebeesthoek. The data were acquired with the Mk\,III VLBI recording
system in Mode B with an effective bandwidth of 28 MHz and correlated
at the Max-Plank Institut f\"ur Radioastronomie in Bonn. No fringes
were found on the baselines to Urumqi. The EVN data were calibrated and
fringe-fitted using the NRAO AIPS package.  The initial amplitude
calibration was accomplished using the system temperature measurements
made during the observations and the a~priori station gain curves.

The snapshot-mode observation of 1156+295 with the VLBA at 5 GHz was
made in June 1996 as part of the VSOP prelaunch VLBA Survey \citep{EF}.
Eight intermediate frequency channels, each 8\,MHz wide, were recorded
for a total bandwidth of 64 MHz. The data were correlated at the NRAO
VLBA correlator (Socorro, NM, USA). The fringe-fitted and calibrated
VLBA data were kindly made available to us by the VSOP/VLBA
pre-launch survey group \citep{EF}.

The post-processing --- including editing, phase and amplitude
self-calibration, imaging and model fitting --- was performed within
AIPS for both the EVN and VLBA data sets.

\section{Results}

Our MERLIN image shows traces of a straight jet at the position angle
of $-20^{\circ}$ and $2''$ in length (Hong et al., in preparation).
This is consistent with the lower-resolution VLA and MERLIN images
reported by \citet{M1}.  There appear to be several regularly spaced
knots within 1 arcsecond of the core, possibly indicating
quasi-periodic activity that has propagated from the base of the jet.

\begin{figure}[hbt]
\centering
\includegraphics[scale=0.38,angle=0]{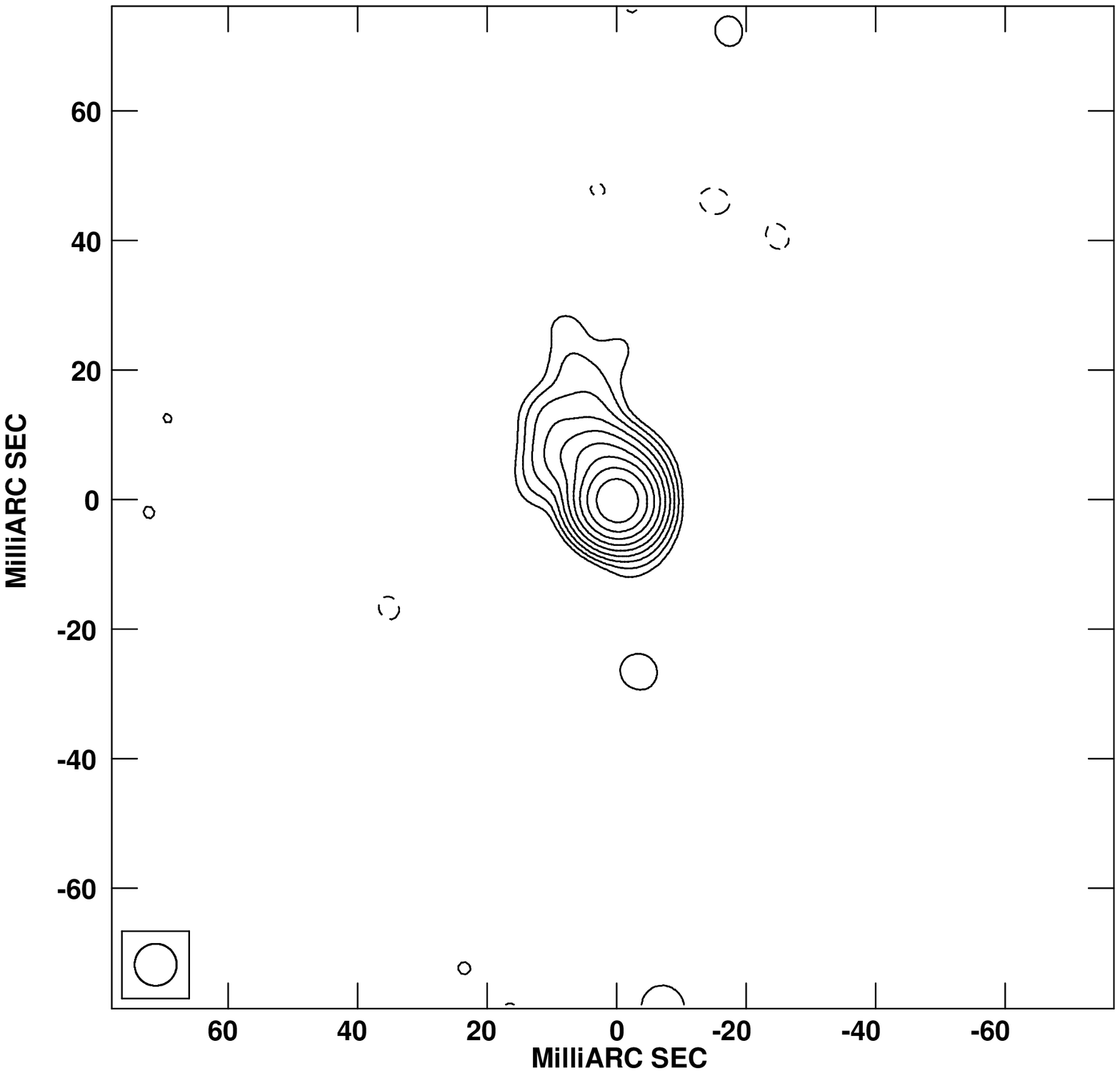}
\includegraphics[scale=0.38,angle=0]{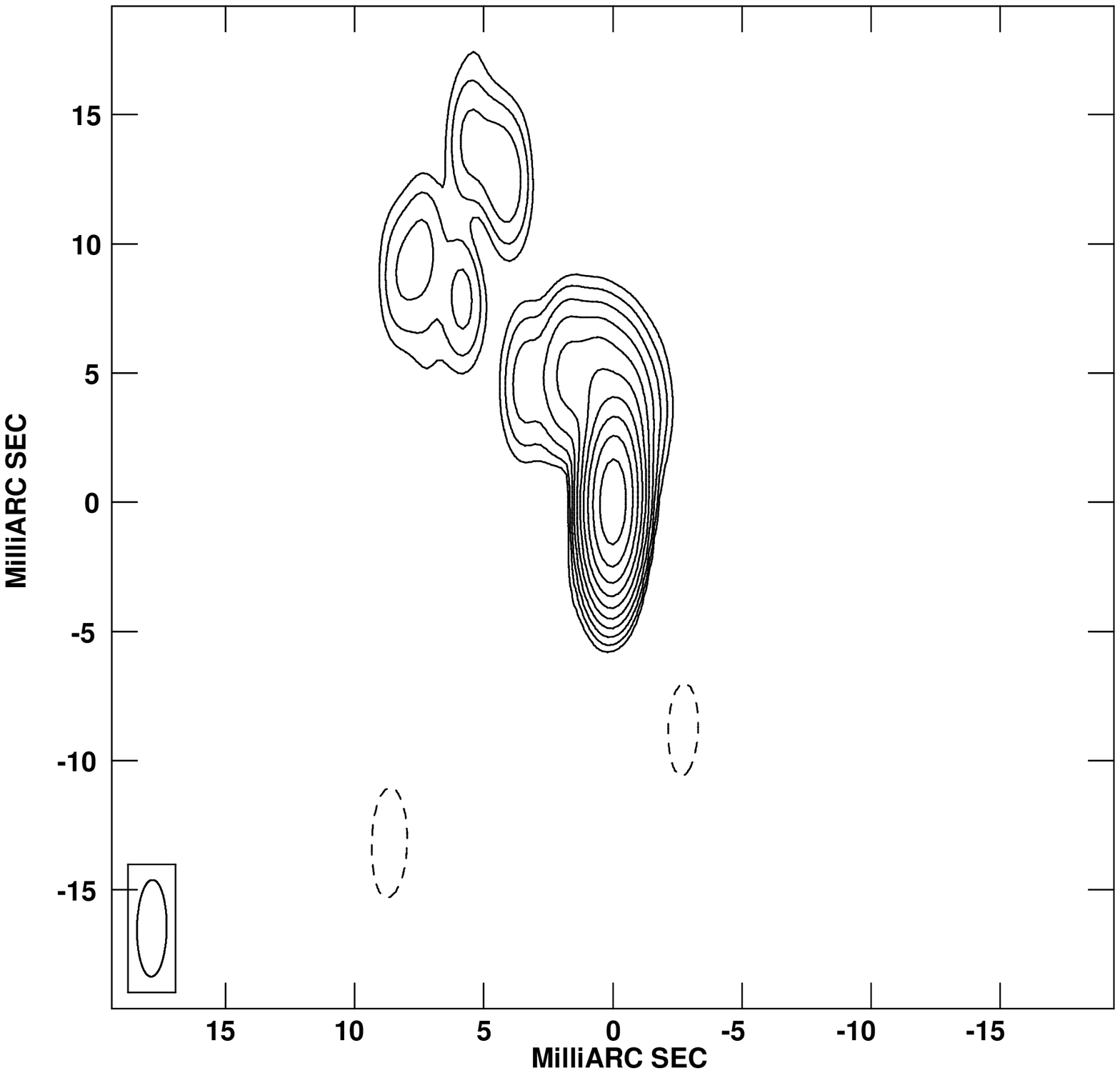}
\caption{The EVN (top) and VLBA (bottom) images of 1156+295 at 5 GHz.
Note the different scales of the two maps.
Contour levels are 8.0$\times 10^{-4}$ ($-$3, 3, 6, 12, 24, 48, 96,
192, 384, 768, 1536). The peak flux densities are 1.18 and 2.08
Jy/beam, respectively.}
\label{fig}
\end{figure}

Because of missing Urumqi data, the u-v coverage of the EVN data has a
big gap between 25 and 90 M$\lambda$. Therefore, we imaged with the
data of the European telescopes first. The image (Fig.~\ref{fig})
reveals a jet that is bending from North to North-East before turning
back to the North, presumably to meet the arcsecond scale jet.

The image of VLBA presents a similar structural pattern
(Fig.~\ref{fig}). It shows more clearly that the jet is bent from the
North to East and then back to the North-East.

A higher resolution EVN image was obtained by using all the data, including
data from Hartebeesthoek and Shanghai (Hong et al., in preparation).
The results show that a new jet component starts at about p.a.
$50^{\circ}$ before bending to the North within 2 milliarcseconds of
the core, which is consistent with the 8 GHz images \citep{P1}. In
Piner and Kingham's images, one can also note that the jet components
start at the position angle of $45^{\circ}$, then move clockwise to the
North before bending to the East sharply.

\section{Discussion}

We note that the peak flux density at two epochs varies from 2.09
Jy/beam for the VLBA map (epoch 1996.43) to 1.2 Jy/beam for the later
EVN map (epoch 1997.14), which is clear evidence of the decrease of the
core brightness (the VLBA synthesized beam of Fig.~\ref{fig} is
smaller than that of EVN, Fig.~\ref{fig}). From the data of the
University of Michigan Radio Astronomy Observatory, the total flux
densities at 14.5, 8.0, and 4.8 GHz have all decreased from mid of 1996
to early 1997 \citep{UMRAO}. In 15\,GHz VLBA maps, the peak flux
density decreased from 1.95 Jy/beam (16 May 1996) to 0.67 Jy/beam (13
March 1997) \citep{K1}. All these indicate  clearly that the flare of
1156+295 was due to the outburst of the central core component. At that
time a new jet component was ejected.

Helical jets have been proposed to explain the bi-modal distribution of
the difference between arcsecond and milliarcsecond structural axes
observed for core-dominated radio sources \citep{C1}. The helical
pattern could result from the precession of the base of the jet (e.g.
\citet{L1}) or fluid-dynamical instabilities in the interaction between
the jet material and surrounding medium \citep{H1}. The structural
oscillations at mas scales can be explained by the orbital motion of a
binary black hole (e.g. \citet{R1} for the case of 1928+738).

The jet of 1156+295 has a structural oscillation on the mas scale and a
large $\Delta$p.a. between its pc-scale and the kpc-scale directions,
which may indicate that the oscillation of the jet is a projection
effect of a helical jet. The observed properties of the jet could be
explained by two effects: the precession of the spin axis of the black
hole emitting the jet and the orbital motion of a binary black hole. In
this case, the jet of 1156+295 could be ejected in the direction very
closely aligned to the line of sight. The jet then would have a high
Doppler  boosting but would demonstrate a relatively low apparent
proper motion. As the direction of jet curves away from the line of
sight, the Doppler factor will decrease (thus, flux density will
decrease also), but the apparent proper motion velocity will increase.
As the bright radio component moves outward, it will reach the viewing
angle of the maximum apparent transverse velocity, while its Doppler
boosting factor will continue to monotonically decrease. This qualitative
model can explain the evolution of the outburst in the core \citep{M1}
and the apparent initial acceleration of the proper motion of the jet
components \citep{P1}. It is also generally consistent with the
$\gamma$-ray observation at the GeV band:  the source demonstrated
several short active periods at the high-energy $\gamma$-ray band but
remained in a quiescent state most of the time between flares
\citep{M3}. The model suggests that the $\gamma$-ray flares may preceed
radio outbursts since the former require higher Doppler boosting.

{\bf Acknowledgements.}
This research was supported by the National Science Foundation and the
Pan Deng Plan of China. XYH thanks JIVE for the hospitality during his
visit in 1998. LIG acknowledges partial support from the European
Comission TMR programme, Access to Large-Scale Facilities under
contract ERBFMGECT950012. The authors are grateful to the staff of EVN,
MERLIN and VLBA for support of the observing projects. The authors
express their gratitude to the team of the VSOP/VLBA pre-launch survey,
particularly Ed Fomalont and Phil Edwards, for the permission to use
their data.  This research has made use of data from the University of
Michigan Radio Astronomy Observatory, which is supported by the
National Science Foundation and by funds from the University of
Michigan. The National Radio Astronomy Observatory is operated by
Associated Universities, Inc. under a Cooperative Agreement with the
National Science Foundation.

\end{document}